\documentclass[9pt,conference]{IEEEtran}
\IEEEoverridecommandlockouts

\usepackage{cite}
\usepackage{amsmath,amssymb,amsfonts}
\usepackage{algorithmic}
\usepackage{graphicx}
\usepackage{textcomp}
\usepackage{booktabs} 
\usepackage{url} 
\usepackage{xcolor}
\def\BibTeX{
{\rm B\kern-.05em{\sc i\kern-.025em b}\kern-.08em
    T\kern-.1667em\lower.7ex\hbox{E}\kern-.125emX}
    }

\makeatletter
\newcommand{\linebreakand}{%
  \end{@IEEEauthorhalign}
  \hfill\mbox{}\par
  \mbox{}\hfill\begin{@IEEEauthorhalign}
}
\makeatother

\begin{document}

\title{CoDiff-VC: A Codec-Assisted Diffusion Model for Zero-shot Voice Conversion
}

\author{\IEEEauthorblockN{1\textsuperscript{st}Yuke Li}
\IEEEauthorblockA{\textit{Northwestern Polytechnical University} \\
Xi'an, China \\
liyk22@mail.nwpu.edu.cn}\\

\IEEEauthorblockN{4\textsuperscript{th}JiXun Yao}
\IEEEauthorblockA{\textit{Northwestern Polytechnical University} \\
Xi'an, China \\
yaojx@mail.nwpu.edu.cn}
\\

\IEEEauthorblockN{7\textsuperscript{th} XiPeng Yang}
\IEEEauthorblockA{\textit{Shanghai Mobvoi Information Technology} \\
Shanghai, China \\
xipeng.yang@mobvoi.com}
\and

\IEEEauthorblockN{2\textsuperscript{nd}Xinfa Zhu}
\IEEEauthorblockA{\textit{Northwestern Polytechnical University} \\
Xi'an, China \\
xfzhu@mail.nwpu.edu.cn}
\\
\IEEEauthorblockN{5\textsuperscript{th} WenJie Tian}
\IEEEauthorblockA{\textit{Northwestern Polytechnical University} \\
Xi'an, China \\
twj@mail.nwpu.edu.cn}
\\
\IEEEauthorblockN{8\textsuperscript{th} Zhifei Li}
\IEEEauthorblockA{\textit{Shanghai Mobvoi Information Technology} \\
Shanghai, China \\
zfli@mobvoi.com}
\and
\IEEEauthorblockN{3\textsuperscript{rd}Hanzhao Li}
\IEEEauthorblockA{\textit{Northwestern Polytechnical University} \\
Xi'an, China \\
lihanzhao.mail@gmail.com}

\\
\IEEEauthorblockN{6\textsuperscript{th} YunLin Chen}
\IEEEauthorblockA{\textit{Shanghai Mobvoi Information Technology} \\
Shanghai, China \\
yunlinchen@mobvoi.com}

\\
\IEEEauthorblockN{9\textsuperscript{th} Lei Xie$^{*}$\thanks{* Corresponding authors.}}
\IEEEauthorblockA{\textit{Northwestern Polytechnical University} \\
Xi'an, China \\
lxie@nwpu.edu.cn}
}
\vspace{10pt}
\maketitle
\vspace{10pt}
\begin{abstract}
Zero-shot voice conversion (VC) aims to convert the original speaker's timbre to any target speaker while keeping the linguistic content. Current mainstream zero-shot voice conversion approaches depend on pre-trained recognition models to disentangle linguistic content and speaker representation. This results in a timbre residue within the decoupled linguistic content and inadequacies in speaker representation modeling. 
In this study, we propose CoDiff-VC, an end-to-end framework for zero-shot voice conversion that integrates a speech codec and a diffusion model to produce high-fidelity waveforms. Our approach involves employing a single-codebook codec to separate linguistic content from the source speech. To enhance content disentanglement, we introduce Mix-Style layer normalization (MSLN) to perturb the original timbre. Additionally, we incorporate a multi-scale speaker timbre modeling approach to ensure timbre consistency and improve voice detail similarity. To improve speech quality and speaker similarity, we introduce dual classifier-free guidance, providing both content and timbre guidance during the generation process. Objective and subjective experiments affirm that CoDiff-VC significantly improves speaker similarity, generating natural and higher-quality speech. 
Audio samples are available on the demo page\footnote{\url{https://aries457.github.io/CoDiff-VC/}}.
\end{abstract}

\begin{IEEEkeywords}
zero-shot voice conversion, multi-scale timbre modeling, diffusion model.
\end{IEEEkeywords}

\section{Introduction}
\begin{figure*}[]
\centering
\includegraphics[width=\linewidth]{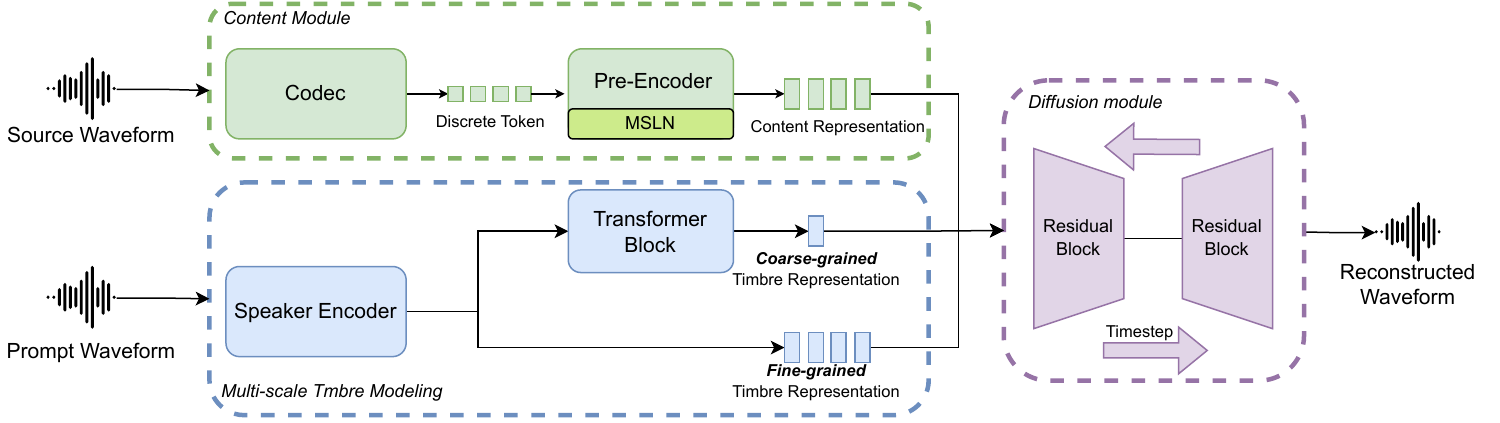}
\caption{Overall architecture of the CoDiff-VC. The content module within the green dashed line is used to extract linguistic content from the source speech, while we employ a multi-scale timbre modeling module to capture the details of speaker timbre. The diffusion module within the purple dashed line reconstructs the speech waveform conditioned on the linguistic content and speaker timbre. }
\vspace{-15pt}
\end{figure*}

Zero-shot voice conversion (VC) aims to transfer the timbre of a source speaker to the timbre of an unseen target speaker while maintaining the original linguistic content~\cite{sisman2020overview}. This approach requires only one utterance from the target speaker, making it applicable in various scenarios like movie dubbing~\cite{yao2023preserving}, speech translation~\cite{sundermann2003vtln}, and speech anonymization~\cite{yoo2020speaker}. In real-world applications, zero-shot voice conversion can enhance personalized voice interactions~\cite{huang2012personalized,csicsman2017transformation} and improve entertainment experiences by converting speaker timbre characteristics. However, since only one target speaker utterance is available, disentangling speech components and simultaneously converting to the target speaker's timbre becomes more challenging. 

    In the zero-shot VC task, there are two primary challenges to address: firstly, how to disentangle the linguistic content and speaker timbre from the source speech; secondly, how to model the speaker representation precisely. To solve the first challenge, many previous approaches~\cite{qian2019autovc,wang2023lm,zhao2018accent,liu2021diffsvc} utilize pre-trained automatic speech recognition (ASR) or self-supervised learning (SSL) models~\cite{baevski2020wav2vec,hsu2021hubert} to extract bottleneck features as linguistic content decoupled from the source speech. Simultaneously, a speaker verification model is employed to extract the speaker representation. The VC model then combines the original linguistic content with the target speaker representation to reconstruct the converted speech. Despite the previous approach achieving some success in zero-shot VC~\cite{DBLP:conf/apsipa/XiaoXYX21,DBLP:journals/corr/abs-2111-03811}, the extracted bottleneck features still contain speaker-related information, leading to poor speaker similarity in the converted speech. Meanwhile, most of the mentioned approaches depend on an acoustic model for predicting the mel-spectrogram-like latent representations and employ a vocoder to reconstruct the representations into speech waveform~\cite{dang2022training,liu2021diffsvc,DBLP:conf/iclr/ShenJ0LL00Z024}. This two-stage framework introduces cascading errors, thereby degrading the quality of the converted speech.

For the second challenge, the previous studies on speaker representation modeling can be broadly divided into two categories: coarse-grained modeling approach~\cite{doddipatla2017speaker,casanova2022yourtts,kinnunen2017non} and fine-grained modeling approach~\cite{zhou2022content,jiang2023mega,lee2022pvae}. In coarse-grained modeling, a pre-trained speaker verification model is utilized to extract a global speaker embedding as the coarse-grained speaker representation. While effective in controlling the overall timbre characteristics of the converted speech, these approaches fall short in capturing detailed speaker timbre information and semantic-related timbre changes, leading to limited speaker similarity between the converted speech and target speaker speech. Conversely, other studies employ an attention mechanism to capture fine-grained speaker representation from multiple reference utterances, with Mega-TTS2~\cite{jiang2023mega} being the most prominent example, which allows for the generation of more natural speech for the target speaker. However, the speaker similarity in converted speech using these approaches relies on the duration of the reference, resulting in a notable degradation in similarity performance when the reference speech duration is excessively short.



In this study, with particular consideration of the above two challenges, we propose CoDiff-VC, a codec-assisted end-to-end framework for zero-shot voice conversion, which can generate high-fidelity waveforms without any auxiliary losses and avoid cascading error issues. 
We employ a pre-trained codec model, featuring only a single codebook, to extract discrete tokens from the source speech as the linguistic content. The single codebook architecture can partially disentangle the speaker's timbre by introducing a speaker reference encoder while retaining accurate linguistic content information. Meanwhile,  we incorporate  Mix-Style layer normalization (MSLN)~\cite{huang2022generspeech} to perturb the timbre information within the discrete tokens, facilitating a more thorough disentanglement of timbre characteristics from the tokens. To improve timbre similarity and consistency, we introduce a multi-scale speaker timbre modeling approach to recover voice details when reconstructing the waveform. Finally, we propose a dual classifier-free guidance strategy to train unconditional models of content and timbre for guiding the reverse process to generate high-quality waveforms.
Objective and subjective experiments demonstrate that CoDiff-VC outperforms the baseline systems in both speech quality and speaker similarity. Ablation studies further demonstrate the effectiveness of each component in our proposed approach.

\section{Proposed Approach}
The overall architecture of our proposed system is shown in Figure 1, which consists of three parts: content module, multi-scale timbre modeling, and diffusion module.
For the content module, we utilize a pre-trained codec model with a reference encoder to encode audio into discrete tokens and employ mix-style layer normalization further to reduce the residual timbre in the discrete tokens. To achieve more comprehensive timbre modeling, we introduce multi-scale speaker timbre modeling, encompassing modeling of both coarse-grained and fine-grained timbre representation. Following this, a denoising diffusion probabilistic model serves as the backbone of the conversion model for an end-to-end reconstruction process. To enhance the quality of the reconstructed speech, we propose a dual classifier-free guidance strategy to effectively guide content and timbre modeling, thereby improving speech quality and speaker similarity.

\subsection{Linguistic Content Extraction}

To disentangle the linguistic content from the source speech, we utilize a pre-trained codec to extract discrete tokens as linguistic content. The architecture closely resembles the codec in ~\cite{tortoise}. Since the majority of information is preserved in the first codebook and lower bit rates make it easier to decouple timbres~\cite{ren2023fewer}, we employ a single-codebook speech codec, comprising only 4096 tokens, to reduce bit rates. And, to disentangle timbre characteristics within the tokens in the codec, we incorporate a reference encoder into the codec. The output of the reference encoder, representing global representation related to acoustic information, is subtracted from the hidden representation of the encoder. This subtraction process imposes implicit constraints on the tokens, leading them to spontaneously disentangle global representations, such as timbre and acoustic features, at lower bit rates.

We use an embedding layer to transform the discrete token into a high-level representation. The token embedding is upsampled to match the target audio's shape, serving as the content condition for the diffusion process. To eliminate the speaker timbre, we incorporate MSLN into the pre-encoder to perturb the original timbre information within the tokens by introducing mismatched timbre information, preventing the generation of timbre-consistent representations.
\subsection{Multi-scale Speaker Modeling}
Many previous studies~\cite{kang2023grad,min2021meta} assume that speaker timbre is a time-invariant global representation, neglecting the potential timbre variations due to intense emotions or specific semantics. In scenarios of zero-shot voice conversion, relying solely on global representation to model speaker timbre may result in a degradation of timbre similarity in the converted speech. To address this limitation and leverage the powerful reconstruction capability of diffusion, we introduce multi-scale speaker timbre modeling within the diffusion model. This approach enables the simultaneous capture of coarse-grained and fine-grained timbre information from reference speech, thereby enhancing the speaker similarity in the converted speech sample. As shown in Figure 1, the speaker timbre modeling module comprises a speaker encoder and a transformer module. Initially, we randomly select an utterance and clip it to a fixed-length segment as the reference speech. The speaker encoder processes this reference to produce frame-level speaker representations, which serve as fine-grained timbre representations. The transformer module then takes these frame-level representations and generates coarse-grained timbre representations through global pooling.
\begin{figure}[t]
  \centering
  \includegraphics[width=160pt]{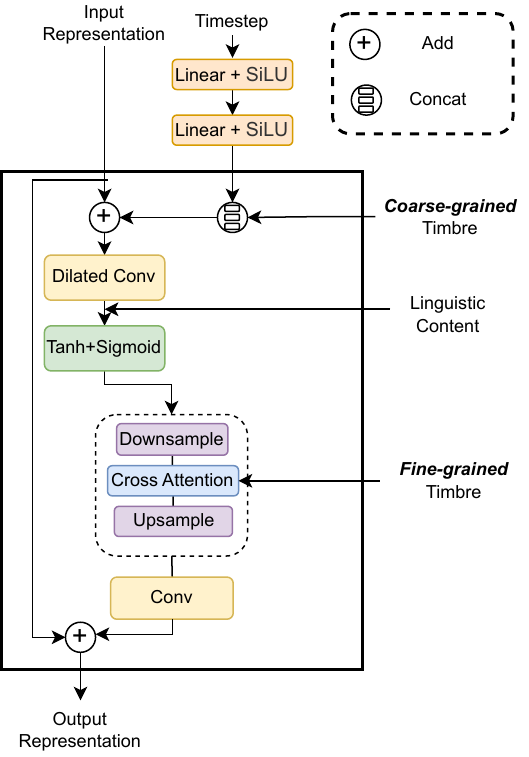}
  \vspace{-8pt}
  \caption{The structure of U-net block.}
  \label{fig:speech_production}
  \vspace{-19pt}
\end{figure}
To integrate timbre characteristics at different levels, we use distinct approaches within the diffusion model. As depicted in Figure 2, the coarse-grained timbre information is concatenated with the timestep embedding and then fed into the residual block. This setup enables the supervision of diffusion by timbre at each timestep when predicting noise. For fine-grained timbre modeling, we introduce a cross-attention convolution module every 4 U-net blocks, associating linguistic content information with timbre. Specifically, the latent representation undergoes 256-time downsampling to match the same dimension as the mel spectrogram to be employed as the query in the cross-attention module, and the fine-grained timbre serves as the keys and values. Ultimately, the upsampling module restores it to the same shape as the audio.

\subsection{Dual Classifier-free Guidance}
To guide the reverse diffusion process, we propose a novel dual classifier-free guidance strategy. We employ classifier-free guidance~\cite{ho2022classifier} to content representation $R_{c}$ and speaker timbre $R_{s}$, respectively. By introducing two distinctive implicit classifiers, both content information and timbre information can guide noise generation in un-condition scenarios. During training, we discard the content condition and timbre condition at a rate of 15\%. The inference process is shown in Eq.(1), where the ${\hat{\epsilon}}_{\theta}$ denotes the denoising network which reverses original sample $x_0$ from Gaussian noise $x_t$. Meanwhile, $w_{c}$ and $w_{s}$ represent the guidance scales of text conditions and timbre conditions for noise prediction, respectively.
\begin{align}
    \hat{\epsilon}_{\theta}\left(x_{t}, t, R_{c},R_{s}\right)\begin{aligned}[t]
    &=(1+w_c+w_s) \cdot \epsilon\left(x_{t}, t, R_{c},R_{s}\right) \\
    &-w_{c} \cdot \epsilon\left(x_{t}, t, R_{s}\right)-w_{s}\cdot \epsilon\left(x_{t}, t,R_{c}\right)
    \end{aligned}
\end{align}
During inference, we need to assign a guidance scale to unconditional estimation, which can determine the impact of real conditional estimation and unconditional estimation on the synthesis process. With each timestep of inference, the coarse-grained information is first modeled before modeling fine-grained information. Therefore, we use an annealing strategy for content representation, where the scale $w_{c}$ gradually decreases over timesteps. Meanwhile, we set the guidance scale $w_{s}$ following a hardening strategy, which gradually increases over timesteps, indicating that fine-grained speaker information is gradually restored to improve speaker similarity.

\subsection{Training and Inference}
We first train the speech codec and then extract quantized discrete tokens for conversion model training. We select audio clips of the same speaker as the reference speech to get a multi-scale timbre representation. Then, with the content representation and timbre representation as condition information, the proposed CoDiff-VC is optimized by minimizing the following unweighted variant of the ELBO~\cite{ho2020denoising} as shown in Eq.(3), which has proved to be effective for high-quality generation~\cite{tortoise}.
\begin{align}
    \bar\alpha _t = \prod_{t=1}^{T}(1-\beta_{t}) 
\end{align}
\begin{align}
    \min _{\theta} L(\theta)=\mathbb{E}_{x_{0}, \epsilon, t}||\epsilon-\epsilon_{\theta}\left(\sqrt{\bar{\alpha}_{t}} x_{0}+\sqrt{1-\bar{\alpha}_{t}} \epsilon, t, \text{cond}\right)||_{2}^{2}
\end{align}
where the $\epsilon \sim N(0,1)$, $\beta_{t}$ is linear noise schedule and $cond$ means condition information. 
For the reverse process, it can also be defined as a Markov chain from the noisy data $x_t$ to the original data $x_0$ as shown in Eq.(4),
where $\mu_{\theta}$ and $\Sigma_{\theta}$ are the mean and variance terms respectively. Finally, in order to ensure the speed and quality of inference, we adopt fast sampling~\cite{chen2020wavegrad} and $T_{\text{infer}}=10$.
\begin{align}
 p_{\theta}\left(x_{t-1} \mid x_{t}\right)=N\left(x_{t-1} ; \mu_{\theta}\left(x_{t},t,\text{cond}\right), \Sigma_{\theta}\left(x_{t}, t\right)\right)
\end{align}

\section{Experiments}
\begin{table*}[!htbp]
\caption{Results of CoDiff-VC compared with baseline systems and Ablation models regarding speaker similarity MOS, naturalness MOS and speech quality 
MOS with confidence intervals of 95\%. SCOS and WER mean speaker cosine similarity and word error rate.}

\label{tab:mos}
\setlength{\tabcolsep}{5mm}
\centering
\renewcommand{\arraystretch}{1.2}
\fontsize{8pt}{8pt}\selectfont
\begin{tabular}{@{}lccc|ccc@{}}
\toprule

Model & Speaker Similarity (MOS↑) & Naturalness (MOS↑) & Speech Quality (MOS↑) & SCOS↑ 
& WER↓ \\ \hline

Ground Truth & - & 4.39$\pm$0.09 & 4.53$\pm$0.12 & - & 2.176\\
YourTTS~\cite{casanova2022yourtts} & 3.43$\pm$0.10 & 3.37$\pm$0.09 & 3.61$\pm$0.11 & 0.574 & 4.850\\
SEF-VC~\cite{li2023sef} & 3.67$\pm$0.12 & 3.77$\pm$0.10 & 3.76$\pm$0.10 & 0.782 & 5.436\\
LM-VC~\cite{wang2023lm} & 3.70$\pm$0.09 & 3.83$\pm$0.11 & 4.03$\pm$0.11 & 0.806 & 5.150\\
Diff-VC~\cite{DBLP:conf/iclr/PopovVGSKW22} & 3.72$\pm$0.09 & 3.65$\pm$0.08 & 4.10$\pm$0.08 & 0.815 & 4.326\\
\midrule
\textbf{CoDiff-VC}& \textbf{3.94$\pm$0.10} & \textbf{4.07$\pm$0.08} & \textbf{4.26$\pm$0.09} & \textbf{0.839} & \textbf{3.121}\\ 
\quad-MSLN & 3.79$\pm$0.10 & 3.95$\pm$0.11 & 4.22$\pm$0.10 & 0.830 & 3.170 \\
\quad-Fine-grained & 3.73$\pm$0.10 & 3.88$\pm$0.10 & 4.20$\pm$0.08 & 0.824 & 3.240\\
\quad-Dual CFG & 3.85$\pm$0.08 & 4.01$\pm$0.11 & 4.10$\pm$0.09 & 0.831 & 3.344\\

\bottomrule
\end{tabular}
\vspace{-10pt}
\end{table*}

\subsection{Datesets and Model Setup} 
We conduct our experiments using the LibriTTS dataset~\cite{zen2019libritts}, which comprises 580 hours of speech from 2400 speakers. To standardize the data, we crop the audio samples to the same length and add silence at the end of shorter samples. The dataset is divided into a training set with speech data from 2390 speakers and a test set with data from the remaining 10 speakers. All audio samples are resampled to 24000 Hz.

In our implementation, we employ a codec based on the Tortoise approach~\cite{tortoise}. We trained the codec model on 4 NVIDIA V100 GPUs with a batch size of 1024 for 300k training steps. For the conversion model, CoDiff-VC adopts a U-net structure with 30 blocks, each having a dimension of 128. It utilizes a linear noise schedule for $\beta_{t} \in\left[1 \times 10^{-4}, 0.02\right]$ with T = 200 in the diffusion model, following the approach in DiffWave~\cite{kong2020diffwave}. CoDiff-VC is trained on 4 NVIDIA V100 GPUs in the training setup with a batch size of 16 for 1M steps, using the Adam optimizer with an initial learning rate of 1e-4. 

\subsection{Baseline Systems}
We adopt YourTTS~\cite{casanova2022yourtts}, SEF-VC~\cite{li2023sef}, LM-VC~\cite{wang2023lm} and Diff-VC ~\cite{DBLP:conf/iclr/PopovVGSKW22} as the baseline systems for comparison with our proposed CoDiff-VC in the zero-shot voice conversion scenario:

\begin{itemize}
    \item \textbf{YourTTS}: A TTS model with a speaker voiceprint model to extract speaker embedding and employ HiFi-GAN vocoder to reconstruct the waveform. The speaker embedding served as a condition of the flow-based decoder and posterior encoder for enhancing the zero-shot multi-speaker generation capability.
    \item \textbf{SEF-VC}: A VC framework with a semantic encoder and a mel encoder. The semantic encoder reconstructs the discrete representations extracted by Hubert into mel spectrograms as the backbone model. The mel encoder processes the mel spectrogram to obtain speaker timbre representation via a powerful position-agnostic cross-attention mechanism.
    \item \textbf{LM-VC}: A two-stage language modeling approach that generates acoustic tokens to recover linguistic content and timbre and reconstructs the fine for acoustic details as converted speech.
    \item \textbf{Diff-VC}: A diffusion-based VC model that employs the average voice encoder to transform mel features into a speaker-average representation in the dataset for the disentanglement problem.
\end{itemize}


\vspace{-5pt}
\subsection{Subjective Evaluation}
We use Mean Opinion Score (MOS) as the subjective metric to evaluate performance between the baseline and proposed systems regarding speaker similarity, naturalness, and speech quality. The test set comprises 20 utterances involving 10 unseen speakers (5 females and 5 males). Using 10 speakers as references, each of the 20 sentences serves as input, generating 200 audio samples. Ten listeners are invited to participate in the subjective evaluation, scoring the similarity between the converted and reference speech. Additionally, Naturalness MOS and Speech Quality MOS assessments are conducted to measure the intelligibility of the synthesized voice.

As shown in Table~\ref{tab:mos}, the subjective evaluation results demonstrate the superior performance of our proposed approach in zero-shot voice conversion. YourTTS solely relies on global timbre representation to model speaker timbre, while SEF-VC utilizes only local timbre representation and the self-supervised model extracting semantic information exhibits residual timbre information, leading to poor timbre similarity. In addition, Diff-VC exhibits lower naturalness MOS than CoDiff-VC, suggesting that our content module extracts better content representations. Notably, these three models are constrained by a two-stage method, first reconstructing the mel and then restoring it through the vocoder, resulting in degradation of generation quality. 
Limited to coarse acoustic tokens, LM-VC faces challenges in restoring unseen speaker timbres. In contrast, CoDiff-VC modeling both coarse and fine-grained timbre, coupled with the application of dual CFG on the diffusion model, enhances the quality of converted speech while effectively capturing timbre information.



\subsection{Objective Evaluation}

To evaluate speaker similarity, we calculate the cosine similarity between the converted speech and the reference speech using a pre-trained ECAPA-TDNN~\cite{desplanques2020ecapa}. For intelligibility evaluation, we calculate the Word Error Rate (WER) for the converted speech. The WER assessment is computed using an open-source Wenet toolkit~\cite{yao2021wenet} trained on the LibriSpeech dataset. 

As shown in Table~\ref{tab:mos}, the objective evaluation results align with the subjective evaluation, showing that CoDiff-VC achieved the highest cosine similarity. SEF-VC, LM-VC, and Diff-VC recorded higher WER scores than CoDiff-VC, suggesting that the discrete representation of content through timbre decoupling in Co-DiffVC facilitates more accurate modeling of content information. Meanwhile, the lower WER results indicate that our proposed CoDiff-VC's converted speech outperforms all baseline intelligibility systems.

\begin{figure}[t]
  \centering
  \includegraphics[width=120pt]{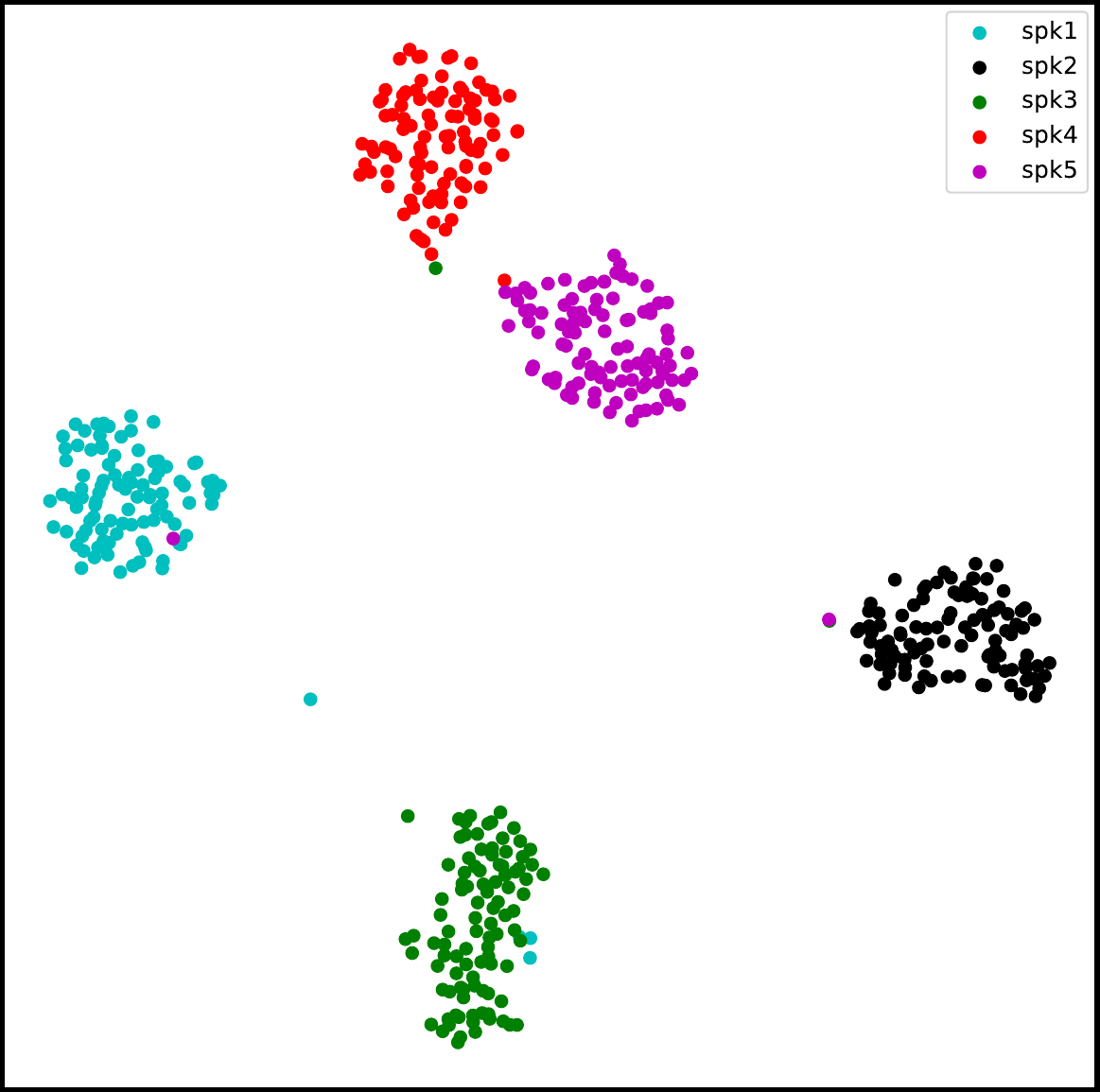}
  \vspace{-5pt}
  \caption{T-SNE visualization of the coarse-grained timbre in multi-speakers settings.}
  \label{fig:tsne}
  \vspace{-15pt}
\end{figure}

To demonstrate the effectiveness of timbre modeling, we visualize the coarse-grained timbre representations through t-SNE~\cite{Maaten2008VisualizingDU}. We select 100 audio samples from five random speakers in the test set and extract coarse-grained timbre representations for clustering using the model’s coarse-grained timbre modeling structure. The clustering results, shown in Figure~\ref{fig:tsne}, indicate that the timbre representations of each speaker are strongly related to the corresponding speakers.

\vspace{-5pt}

\subsection{Ablation Study}

To evaluate the effectiveness of each component in CoDiff-VC, we conduct an ablation study to analyze the impact of three key components: MSLN, multi-scale timbre modeling, and dual classifier-free guidance. We compare CoDiff-VC with three variants: first, without the MSLN module (-MSLN); second, using only coarse-grained speaker timbre instead of multi-scale ones (-fine-grained); and third, without dual classifier-free guidance (-dual cfg), utilizing only classifier-free guidance for content representation.

As illustrated in Table~\ref{tab:mos}, the speaker similarity and speech naturalness are significantly degraded when removing the MSLN module. These results suggest that the MSLN module is crucial for retaining content information while eliminating residual timbre information, playing an important role in disentangling linguistic content. In terms of timbre modeling, relying solely on single-scale speaker representation results in a notable decline in speaker similarity, indicating the critical role of the multi-scale timbre modeling module in capturing the details of speaker timbre. Additionally, naturalness experiences degradation with only coarse-grained timbre modeling, highlighting the importance of timbral details in synthesizing natural speech. Furthermore, the model without dual classifier-free guidance exhibits lower speaker similarity and speech quality, emphasizing the contribution of additional regular timbre information guidance to improved timbre modeling and speech quality.

\vspace{-5pt}
\section{Conclusions}
In this paper, we propose CoDiff-VC, a codec-assisted end-to-end framework designed for zero-shot voice conversion. We employ a pre-trained codec and a pre-encoder to decouple discrete content representations from the source speech. Meanwhile, we introduce a multi-scale speaker timbre modeling module within the diffusion model for modeling fine-grained timbre details. Our proposed CoDiff-VC excels in generating converted audio for unseen target speakers with more similar voice characteristics, higher naturalness, and improved speech quality. Additionally, we present a dual classifier-free guidance method, creating implicit classifiers for content and speaker representations to enhance the inference results of the diffusion model. However, it's worth noting that the inference speed of diffusion remains relatively slow. We aim to address and improve the reverse process of diffusion in our future work.

\bibliographystyle{IEEEbib}
\bibliography{refs}

\begin{thebibliography}{10}

\bibitem{sisman2020overview}
Berrak Sisman, Junichi Yamagishi, Simon King, and Haizhou Li,
\newblock ``An overview of voice conversion and its challenges: From statistical modeling to deep learning,''
\newblock {\em IEEE/ACM Trans. Audio Speech Lang. Process.}, vol. 29, pp. 132--157, 2020.

\bibitem{yao2023preserving}
Jixun Yao, Yi~Lei, Qing Wang, Pengcheng Guo, Ziqian Ning, Lei Xie, Hai Li, Junhui Liu, and Danming Xie,
\newblock ``Preserving background sound in noise-robust voice conversion via multi-task learning,''
\newblock in {\em Proc. ICASSP}, 2023, pp. 1--5.

\bibitem{sundermann2003vtln}
David S{\"u}ndermann, Hermann Ney, and Harald H{\"o}ge,
\newblock ``Vtln-based cross-language voice conversion,''
\newblock {\em Proc. ASRU}, pp. 676--681, 2003.

\bibitem{yoo2020speaker}
In-Chul Yoo, Keonnyeong Lee, Seong-Gyun Leem, Hyunwoo Oh, Bonggu Ko, and Dongsuk Yook,
\newblock ``Speaker anonymization for personal information protection using voice conversion techniques,''
\newblock {\em IEEE Access}, vol. 8, pp. 198637--198645, 2020.

\bibitem{huang2012personalized}
Yi-Chin Huang, Chung-Hsien Wu, and Yu-Ting Chao,
\newblock ``Personalized spectral and prosody conversion using frame-based codeword distribution and adaptive crf,''
\newblock {\em IEEE/ACM Trans. Audio Speech Lang. Process.}, vol. 21, pp. 51--62, 2013.

\bibitem{csicsman2017transformation}
Berrak Sisman, Haizhou Li, and Kay~Chen Tan,
\newblock ``Transformation of prosody in voice conversion,''
\newblock {\em Proc. {APSIPA ASC}}, pp. 1537--1546, 2017.

\bibitem{qian2019autovc}
Kaizhi Qian, Yang Zhang, Shiyu Chang, Xuesong Yang, and Mark~A. Hasegawa-Johnson,
\newblock ``Autovc: Zero-shot voice style transfer with only autoencoder loss,''
\newblock {\em Proc. ICML}, pp. 5210--5219, 2019.

\bibitem{wang2023lm}
Zhichao Wang, Yuan-Jui Chen, Linfu Xie, Qiao Tian, and Yuping Wang,
\newblock ``{LM-VC}: Zero-shot voice conversion via speech generation based on language models,''
\newblock {\em IEEE Sig. Proc. Lett.}, pp. 1157--1161, 2023.

\bibitem{zhao2018accent}
Guanlong Zhao, Sinem Sonsaat, John~M. Levis, Evgeny Chukharev-Hudilainen, and Ricardo Gutierrez-Osuna,
\newblock ``Accent conversion using phonetic posteriorgrams,''
\newblock {\em Proc. {ICASSP}}, pp. 5314--5318, 2018.

\bibitem{liu2021diffsvc}
Songxiang Liu, Yuewen Cao, Dan Su, and Helen Meng,
\newblock ``Diffsvc: A diffusion probabilistic model for singing voice conversion,''
\newblock in {\em Proc. {ASRU}}. IEEE, 2021, pp. 741--748.

\bibitem{baevski2020wav2vec}
Alexei Baevski, Yuhao Zhou, Abdelrahman Mohamed, and Michael Auli,
\newblock ``Wav2vec 2.0: A framework for self-supervised learning of speech representations,''
\newblock {\em Proc. {NIPS}}, vol. 33, pp. 12449--12460, 2020.

\bibitem{hsu2021hubert}
Wei-Ning Hsu, Benjamin Bolte, Yao-Hung~Hubert Tsai, Kushal Lakhotia, Ruslan Salakhutdinov, and Abdel rahman Mohamed,
\newblock ``{HuBERT}: Self-supervised speech representation learning by masked prediction of hidden units,''
\newblock {\em IEEE/ACM Trans. Audio Speech Lang. Process.}, vol. 29, pp. 3451--3460, 2021.

\bibitem{DBLP:conf/apsipa/XiaoXYX21}
Ruitong Xiao, Xiaofen Xing, Jichen Yang, and Xiangmin Xu,
\newblock ``{CA-VC}: A novel zero-shot voice conversion method with channel attention,''
\newblock {\em Proc. {APSIPA ASC}}, pp. 800--807, 2021.

\bibitem{DBLP:journals/corr/abs-2111-03811}
Haozhe Zhang, Zexin Cai, Xiaoyi Qin, and Ming Li,
\newblock ``{SIG-VC}: A speaker information guided zero-shot voice conversion system for both human beings and machines,''
\newblock {\em Proc. {ICASSP}}, pp. 6567--65571, 2021.

\bibitem{dang2022training}
Trung Dang, Dung~T. Tran, Peter Chin, and Kazuhito Koishida,
\newblock ``Training robust zero-shot voice conversion models with self-supervised features,''
\newblock {\em Proc. {ICASSP}}, pp. 6557--6561, 2022.

\bibitem{DBLP:conf/iclr/ShenJ0LL00Z024}
Kai Shen, Zeqian Ju, Xu~Tan, Eric Liu, Yichong Leng, Lei He, Tao Qin, Sheng Zhao, and Jiang Bian,
\newblock ``Naturalspeech 2: Latent diffusion models are natural and zero-shot speech and singing synthesizers,''
\newblock in {\em Proc. {ICLR}}, 2024.

\bibitem{doddipatla2017speaker}
Rama~Sanand Doddipatla, Norbert Braunschweiler, and Ranniery Maia,
\newblock ``Speaker adaptation in dnn-based speech synthesis using d-vectors,''
\newblock in {\em Proc. {INTERSPEECH}}, 2017, pp. 3404--3408.

\bibitem{casanova2022yourtts}
Edresson Casanova, Julian Weber, Christopher~Dane Shulby, Arnaldo~C{\^a}ndido J{\'u}nior, Eren G{\"o}lge, and Moacir~Antonelli Ponti,
\newblock ``{YourTTS}: Towards zero-shot multi-speaker tts and zero-shot voice conversion for everyone,''
\newblock in {\em Proc. ICML}, 2021.

\bibitem{kinnunen2017non}
Tomi~H. Kinnunen, Lauri Juvela, Paavo Alku, and Junichi Yamagishi,
\newblock ``Non-parallel voice conversion using i-vector plda: towards unifying speaker verification and transformation,''
\newblock {\em Proc. {ICASSP}}, pp. 5535--5539, 2017.

\bibitem{zhou2022content}
Yixuan Zhou, Changhe Song, Xiang Li, Luwen Zhang, Zhiyong Wu, Yanyao Bian, Dan Su, and Helen Meng,
\newblock ``Content-dependent fine-grained speaker embedding for zero-shot speaker adaptation in text-to-speech synthesis,''
\newblock in {\em Proc. {INTERSPEECH}}, 2022, pp. 2573--2577.

\bibitem{jiang2023mega}
Ziyue Jiang, Jinglin Liu, Yi~Ren, Jinzheng He, Chen Zhang, Zhe Ye, Pengfei Wei, Chunfeng Wang, Xiang Yin, Zejun Ma, and Zhou Zhao,
\newblock ``{Mega-TTS} 2: Zero-shot text-to-speech with arbitrary length speech prompts,''
\newblock {\em ArXiv}, vol. abs/2307.07218, 2023.

\bibitem{lee2022pvae}
Ji-Hyun Lee, Sang-Hoon Lee, Ji-Hoon Kim, and Seong-Whan Lee,
\newblock ``{PVAE-TTS}: Adaptive text-to-speech via progressive style adaptation,''
\newblock {\em Proc. {ICASSP}}, pp. 6312--6316, 2022.

\bibitem{huang2022generspeech}
Rongjie Huang, Yi~Ren, Jinglin Liu, Chenye Cui, and Zhou Zhao,
\newblock ``Generspeech: Towards style transfer for generalizable out-of-domain text-to-speech,''
\newblock {\em Proc. {NIPS}}, vol. 35, pp. 10970--10983, 2022.

\bibitem{tortoise}
James Betker,
\newblock ``Better speech synthesis through scaling,''
\newblock {\em ArXiv}, vol. abs/2305.07243, 2023.

\bibitem{ren2023fewer}
Yong Ren, Tao Wang, Jiangyan Yi, Le~Xu, Jianhua Tao, Chuyuan Zhang, and Jun Zhou,
\newblock ``Fewer-token neural speech codec with time-invariant codes,''
\newblock {\em ArXiv}, vol. abs/2310.00014, 2023.

\bibitem{kang2023grad}
Minki Kang, Dong Min, and Sung~Ju Hwang,
\newblock ``Grad-stylespeech: Any-speaker adaptive text-to-speech synthesis with diffusion models,''
\newblock {\em Proc. {ICASSP}}, pp. 1--5, 2023.

\bibitem{min2021meta}
Dongchan Min, Dong~Bok Lee, Eunho Yang, and Sung~Ju Hwang,
\newblock ``Meta-stylespeech: Multi-speaker adaptive text-to-speech generation,''
\newblock in {\em Proc. ICML}, 2021, pp. 7748--7759.

\bibitem{ho2022classifier}
Jonathan Ho,
\newblock ``Classifier-free diffusion guidance,''
\newblock {\em ArXiv}, vol. abs/2207.12598, 2022.

\bibitem{ho2020denoising}
Jonathan Ho, Ajay Jain, and Pieter Abbeel,
\newblock ``Denoising diffusion probabilistic models,''
\newblock {\em Proc. {NIPS}}, vol. 33, pp. 6840--6851, 2020.

\bibitem{chen2020wavegrad}
Nanxin Chen, Yu~Zhang, Heiga Zen, Ron~J. Weiss, Mohammad Norouzi, and William Chan,
\newblock ``{WaveGrad}: Estimating gradients for waveform generation,''
\newblock in {\em Proc. {ICLR}}, 2021.

\bibitem{li2023sef}
Junjie Li, Yiwei Guo, Xie Chen, and Kai Yu,
\newblock ``{SEF-VC}: Speaker embedding free zero-shot voice conversion with cross attention,''
\newblock {\em ArXiv}, vol. abs/2312.08676, 2023.

\bibitem{DBLP:conf/iclr/PopovVGSKW22}
Vadim Popov, Ivan Vovk, Vladimir Gogoryan, Tasnima Sadekova, Mikhail~Sergeevich Kudinov, and Jiansheng Wei,
\newblock ``Diffusion-based voice conversion with fast maximum likelihood sampling scheme,''
\newblock in {\em Proc, ICLR}, 2022.

\bibitem{zen2019libritts}
Heiga Zen, Viet-Trung Dang, Robert A.~J. Clark, Yu~Zhang, Ron~J. Weiss, Ye~Jia, Z.~Chen, and Yonghui Wu,
\newblock ``Libritts: A corpus derived from librispeech for text-to-speech,''
\newblock in {\em Proc. {INTERSPEECH}}, 2019, pp. 1526--1530.

\bibitem{kong2020diffwave}
Zhifeng Kong, Wei Ping, Jiaji Huang, Kexin Zhao, and Bryan Catanzaro,
\newblock ``{DiffWave}: A versatile diffusion model for audio synthesis,''
\newblock {\em Proc. {ICLR}}, pp. 562--571, 2020.

\bibitem{desplanques2020ecapa}
Brecht Desplanques, Jenthe Thienpondt, and Kris Demuynck,
\newblock ``Ecapa-tdnn: Emphasized channel attention, propagation and aggregation in tdnn based speaker verification,''
\newblock in {\em Proc. {INTERSPEECH}}, 2020, pp. 3830--3834.

\bibitem{yao2021wenet}
Zhuoyuan Yao, Di~Wu, Xiong Wang, Binbin Zhang, Fan Yu, Chao Yang, Zhendong Peng, Xiaoyu Chen, Lei Xie, and Xin Lei,
\newblock ``{WeNet}: Production oriented streaming and non-streaming end-to-end speech recognition toolkit,''
\newblock in {\em Proc. {INTERSPEECH}}, 2021, pp. 4054--4058.

\bibitem{Maaten2008VisualizingDU}
Laurens van~der Maaten and Geoffrey~E. Hinton,
\newblock ``Visualizing data using t-sne,''
\newblock 2008, vol.~9, pp. 2579--2605.

\end{thebibliography}

\end{document}